\documentclass[12pt]{article}
\usepackage{amsmath,amssymb,amsthm,amsxtra,overpic,bbm,bm,epsfig,subfigure}
\usepackage[colorlinks=true,citecolor=blue,linkcolor=blue]{hyperref}
\usepackage{mathrsfs}
\usepackage{enumitem}
\usepackage{graphicx}
\usepackage{color}
\usepackage{comment}
\usepackage{float}
\usepackage{cite}
\usepackage{multirow}
\usepackage{soul}
\usepackage{slashed,stmaryrd}
\usepackage{lineno}
\usepackage{xspace}
\usepackage{ulem}

\def\thefootnote{\fnsymbol{footnote}}

\def\tabnotefont{\fontsize{9}{10}\selectfont}%

\begin{document}

\vspace{0.2cm}

\begin{center}
{\Large\bf A Methanol-mediated Room-Temperature Synthesis of Tellurium-Loaded Liquid Scintillators for Neutrinoless Double Beta Decay Search}
\end{center}

\vspace{0.2cm}

\begin{center}
{\bf Ya-Yun Ding $^{a}$}\footnote{E-mail: dingyy@ihep.ac.cn},
\quad
{\bf Meng-Chao Liu $^{a}$},
\quad
{\bf Gao-Song Li $^{a, b}$}\footnote{E-mail: ligs@ihep.ac.cn},
\quad
{\bf Liang-Jian Wen $^{a, b}$\footnote{E-mail: wenlj@ihep.ac.cn},
\quad
{\bf Fei Liu $^{a}$}},
\quad
{\bf Feng Liu $^{a}$},
\quad
{\bf Jia-Yu Jiang $^{a}$}, 
\quad
{\bf Zhi-Qi Zhang $^{d}$},
\quad
{\bf Wen-Jie Li $^{a}$},
\quad
{\bf Zhi-Yong Zhang $^{a,c}$},
\quad
\\
{\small $^a$Institute of High Energy Physics, Chinese Academy of Sciences, Beijing 100049, China}\\
{\small $^b$State Key Laboratory of Particle Detection and Electronics (Institute of High Energy Physics, CAS), Beijing 100049, China}\\
{\small $^c$University of Chinese Academy of Sciences, Beijing 100049, China}\\
{\small $^d$Beijing Normal University, Beijing 100875, China}
\end{center}

\vspace{1.5cm}

\begin{abstract}
This study establishes a methanol-mediated room-temperature synthesis approach for tellurium-diol (Te-diol) compounds for use in tellurium-loaded liquid scintillator (Te-LS). The synthesis involves the direct reaction of telluric acid with diols (e.g., 1,2-hexanediol) in methanol (MeOH) under ambient conditions (25$\pm$5\textdegree C), with the key features of lower energy consumption and enhanced safety compared with high-temperature azeotropic distillation method. Mechanistic studies reveal that MeOH serves not merely as a solvent but also exhibits a catalytic effect, playing a dual role in this water-free, heterogeneous room-temperature synthesis. The Te-diol compounds enable fabrication of high-performance Te samples exhibiting exceptional optical transparency (attenuation length = 20.1$\pm$1.1 m at $\lambda$=430 nm for 1\% Te mass loading), which is reported here for the first time. Furthermore, the Te-LS achieves long-term spectral stability approaching or exceeding one year for both 1\% and 3\% Te mass loadings, and demonstrates a light yield comparable those of both the azeotropic distillation method and the SNO+ collaboration's Type I loading method, albeit modestly lower than that of their Type II method. The developed protocol offers the potential for a more energy efficient alternative for large-scale Te-LS production, particularly valuable for next-generation neutrinoless double-beta decay experiments.
\end{abstract}

\begin{flushleft}
\hspace{0.9cm} Keywords: tellurium-loaded liquid scintillator, Te-diol, neutrinoless double beta decay

\end{flushleft}

\def\thefootnote{\arabic{footnote}}
\setcounter{footnote}{0}

\newpage

%%%%%%%%%%%%%%%%%%%%%%%%%%%%%%%
\section{Introduction}
\label{sec:introduction}

The search for neutrinoless double beta decay  ($0\nu\beta\beta$) represents one of the most compelling frontiers in particle physics, with profound implications for our understanding of neutrino properties and the fundamental symmetries of nature. If observed, $0\nu\beta\beta$ would confirm the Majorana nature of neutrinos, provide a measurement of the absolute neutrino mass scale, and potentially shed light on the matter-antimatter asymmetry in the universe~\cite{Fukugita:1986hr,Deppisch:2017ecm,Buchmuller:2005eh}. Given its significance, numerous experimental efforts worldwide have pursued this rare decay using a variety of techniques, including high-purity germanium detectors~\cite{GERDA:2018pmc,Majorana:2024,LEGEND:2025}, cryogenic bolometers~\cite{CUORE:2022,CUPID:2024}, liquid xenon time-projection chambers~\cite{EXO-200:2019rkq,nEXO:2018}, and large liquid scintillator detectors \cite{kz:2023}.

Among the emerging approaches, tellurium-loaded liquid scintillator (Te-LS) has gained attention as a promising medium for $0\nu\beta\beta$ searches. The Te-LS approach is adopted by the SNO+ experiment~\cite{SNO+:2015,SNO:2021xpa}, and is also considered by the proposed THEIA~\cite{THEIA:2020}. The JUNO experiment, featuring the world largest 20-kton LS detector, is recently online~\cite{juno_ppnp}. It has great potential to upgrade for $0\nu\beta\beta$ search~\cite{juno_0vbb_cpc}. The Te-LS technique leverages the advantages of liquid scintillator (LS) - large target masses, low energy thresholds, and scalability - while incorporating tellurium-130 (Te-130), a candidate isotope with a high natural abundance (34\%) and a favorable Q-value (2527 keV). This combines efficient background suppression through pulse-shape discrimination and spatial resolution with the potential for ultra-low radioactive purity. Moreover, the dissolved tellurium (Te) in organic LS enables homogeneous loading without compromising optical transparency, offering a competitive balance between sensitivity and technical feasibility.
0{$\nu$}{$\beta$}{$\beta$} detection represents an ultra-low-background and ultra-high-sensitivity measurement, imposing stringent requirements on the optical performance and long-term stability of Te-LS. Various technical approaches have been explored for incorporating Te into organic LS, including direct dissolution with the aid of surfactants or dispersion of Te and its compounds as nanoparticles. However, these methods often suffer from insufficient long-term stability. Based on the concept of solvation dynamics and structural compatibility, the optimal strategy involves synthesizing organotellurium compounds designed to mimic the solvent's physicochemical properties, ensuring the formation of homogeneous and stable true solutions in LS.

Several reports have been published on the synthesis of organotellurium compounds and their application in the preparation of Te-LS\cite{SNO+:2015,SNO+:2017,SNO:2021xpa,IHEP:2023,SNO+:2023,Suslov:2023vho,Suslov:2022fos}. Based on the reactants used with inorganic tellurium compounds, these studies can be categorized into two groups. The first involves synthesis using elemental tellurium and 2-methylvaleric acid (pivalic acid, or isovaleric acid) by a Russian research group\cite{Suslov:2023vho,Suslov:2022fos}, providing a different approach compared to other research groups. The second category, which comprises the majority of the research, focuses on synthesis using telluric acid (TeA) and diol compounds~\cite{SNO+:2017,SNO:2021xpa,IHEP:2023,SNO+:2023}. This line of inquiry began with the SNO+ collaboration. They were the first to employ aqueous phase synthesis techniques for incorporating Te into liquid scintillators and laid the foundation for the basic principles of diol loading. In 2017, by selecting TeA (in aqueous solution) to react with 1,2-butanediol (BD), they synthesized Te-BD compounds, obtaining Te-LS with a high light yield ($\sim$65\% of pure scintillator at 0.5\% Te loading)\cite{SNO+:2017}. However, a few samples have exhibited crystallisation behaviour indicative of exposure to water vapour. In 2021, the SNO+ group reported the use of N,N-dimethyldodecylamin(DDA) as a stabilizing agent for the TeBD-LAB cocktail as well as an additive to increase the light yield of the Te-LS~\cite{SNO:2021xpa}. In early 2023, we reported an azeotropic distillation method~\cite{IHEP:2023} to synthesize Te-diol compounds for Te-LS. This method is primarily characterized by a "water-free" reaction system. It utilizes solid TeA directly as a reactant with 1,2-hexanediol (HD) without introducing any additional water into the system. Furthermore, azeotropic distillation is employed to continuously remove the water produced during the reaction, thereby promoting the formation of Te-diol compounds and enhancing product stability. The Te-HD product (without DDA) exhibits extremely high solubility in linear alkylbenzene (LAB), and DDA was found to increase the reaction rate between TeA and the diol. This approach yields optically transparent and long-term stable Te-LS formulations, with a light yield of $\sim$55\% of pure scintillator at 0.5\% Te loading. Later in 2023, the SNO+ group reported two types of Te loading methods~\cite{SNO+:2023}. Both approaches result in a compound that is completely miscible with LAB scintillator with excellent optical transparency and loading stability verified to be at least in excess of 5.5 years at room temperature. The Type I loading method builds upon the 2017 aqueous synthesis approach, investigating DDA as a crucial solubilizer and stabilizer. Furthermore, the molar ratio of DDA to Te was experimentally optimized to a range of 0.25–0.75 based on light yield measurements. Type II loading employs DDA, TeA, and BD as starting reactants without adding water to dissolve TeA. When the entire process is conducted in LAB without applying heat, the Te-LS prepared from this product shows significantly reduced fluorescence quenching. Characterizations utilizing Electrospray Ionisation (ESI) Mass Spectrometry (MS) and NMR spectroscopy demonstrate a clear difference between the two methods: the Type II product is devoid of polymers and is composed almost entirely of tellurium monomers with two bidentate BD attachments. This monomeric nature is identified as the key factor responsible for suppressing the fluorescence quenching. Nevertheless, the by-product water in Type II loading must still be removed. Water removal without heating can be achieved via techniques like nitrogen sparging. However, applying these methods to large-scale production poses practical challenges, particularly regarding the removal of the vast majority of water from the product within a short timeframe.

Therefore, in the reaction system of TeA and diol, utilizing an organic solvent that is easily removed via distillation to assist synthesis remains a viable pathway with its own advantages. Although the azeotropic distillation method significantly improves product stability, the process requires acetonitrile as a solvent and high-temperature reactions, followed by the complete distillation of the low-boiling-point acetonitrile—a step that poses safety risks during large-scale production due to the flammability and volatility of the solvent. To address this limitation, this work presents a safer, room-temperature synthesis route for producing Te-diol compounds compared with the azeotropic distillation method. This method retains the water-free reaction system and eliminates the need for distilling hazardous solvents while preserving the critical optical properties and stability of the resulting Te-LS. In this methanol-mediated approach, MeOH not only serves as the solvent but also exhibits a catalytic effect, facilitating a rapid reaction at room temperature. Although MeOH also has a low boiling point, both the synthesis and the removal of MeOH can be performed at room temperature, which mitigates the safety risks to a certain extent. Admittedly, compared to the aqueous synthesis reported in Ref. ~\cite{SNO+:2023}, the use of MeOH still poses substantial safety risks. Nevertheless, exploring water free synthesis routes remains highly valuable. The room-temperature synthesis method presented herein demonstrates a potential scalable pathway suitable for mass production, offering an alternative route for the practical application of Te-LS in next-generation neutrinoless double-beta decay experiments.

\section{Establishment of Room-Temperature Synthesis approach}
\label{sec:synthesis method}

The reaction between TeA and diols is reversible, following the equilibrium:
\\\centerline{Te(OH)$_6$ + diol $\leftrightarrows$ Te-diol + H$_2$O}
\\The removal of the byproduct water shifts the equilibrium toward Te-diol formation, thereby enhancing reaction yield and reducing required reaction time. Azeotropic distillation constitutes one established technique for water removal; in contrast, a methanol-mediated reaction process, conducted at room temperature, has been explored as an alternative methodology.

\subsection{Comparative Evaluation of Solvent Effects on Reaction Kinetics}
\label{subsubsec:solvent}

Reference ~\cite{IHEP:2023} reports the synthesis of Te-HD via the reaction of HD with TeA using azeotropic distillation to prepare Te-LS. As the method presented herein is an improvement upon this technique, we have retained HD as the primary subject to enable a direct comparison of the synthesized products. To achieve the reaction between TeA and HD under relatively mild conditions, such as at room temperature, various solvents were investigated.

All reactions were catalyzed by DDA to enable rapid kinetic assessment with fixed stoichiometry (n$_\mathrm{Te}$:n$_\mathrm{HD}$:n$_\mathrm{DDA}$ = 1:3:0.5). Identical masses of TeA, HD, and DDA were used for each synthesis, with a constant volume of solvent employed in every case. The reactions were conducted in stoppered glass vessels; the mixtures were shaken once daily and left undisturbed for the remainder of the time. The reaction time was defined as the duration required for the complete disappearance of TeA, serving as a metric to compare reaction rates across different solvents. The results indicate that under these experimental conditions, the reaction in MeOH was the fastest, reaching completion in approximately 4 days. Reactions in toluene, acetone, dichloromethane, and pseudocumene required approximately 7 days; m-xylene took about 9 days; acetonitrile about 10 days; trichloromethane about 14 days; and cyclohexane and hexane about 17 days. In LAB, the reaction took approximately 23 days and resulted in phase separation. In the absence of a solvent, significant TeA residue remained after 23 days. Given that these reaction times already demonstrated distinct solvent dependence and significant variations, no further experiments with more precisely controlled conditions were conducted.

The acceleration mechanism of MeOH is discussed in Sec.~\ref{subsubsec:Mechanistic Study}. For other solvents, the reaction rate shows a positive correlation with both polarity and water solubility - solvents with higher polarity and better water miscibility generally demonstrate superior reaction rates. Non-polar solvents with poor water solubility, such as cyclohexane and hexane, exhibit significantly slower reaction rates for the TeA-HD system, though still faster than the solvent-free condition. This enhancement can be attributed to the dilution effect of the solvent, which reduces the concentration of reaction products (water and Te-diol compounds) and consequently drives the reversible reaction toward product formation according to Le Chatelier's principle.

Ref.~\cite{SNO+:2023}  reports a cold synthesis (type II laoding) method for the reaction of BD with TeA and DDA in a LAB medium. However, in TeA-HD reaction system using LAB as solvent, phase separation was observed after room-temperature reaction despite complete consumption of TeA. The volumes of the upper and bottom phases are approximately the same. X-ray fluorescence analysis revealed that both layers contained Te, and the upper layer showed significantly lower Te content compared to the bottom layer. A similar phase separation was observed when substituting BD for HD. Considering the description in Ref. ~\cite{SNO+:2023}, this phenomenon is likely related to the removal of water upon completion of the reaction. The experimental results also validate this point. Therefore, it is necessary to remove the water generated as a by-product prior to the preparation of Te-LS.

The kinetic studies demonstrated that MeOH exhibited the highest reaction rate among the tested solvents. As the simplest monohydric alcohol, MeOH was subsequently compared with two structurally similar alcohols - ethanol and 1-propanol - as reaction solvents under identical experimental conditions. The reaction proceeded fastest when MeOH was used as solvent. Both ethanol and 1-propanol showed significantly slower reaction rates compared to MeOH. Therefore, MeOH was chosen due to its rapid reaction rate under water-free conditions. Despite potential safety considerations, the ability to conduct the synthesis and vacuum distillation at ambient temperature makes this approach a valuable subject for detailed study.

\subsection{Optimization of MeOH Dosage}
\label{subsubsec:MeOH dosage}
\begin{figure}[!htb]
\centering
\includegraphics[width=0.45\columnwidth]{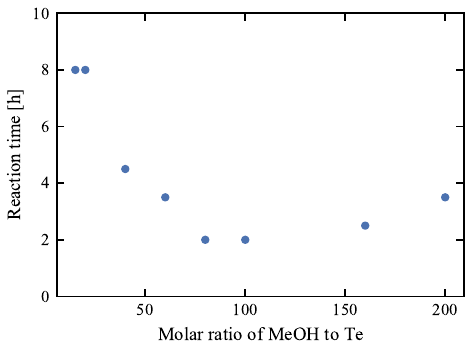}
\caption{Correlation between MeOH dosage and reaction time. } \label{fig:MeOH vs time}
\end{figure}

To evaluate the influence of MeOH dosage on reaction efficiency, a series of experiments were conducted with varying MeOH-to-Te ratios while maintaining other conditions constant (n$_\mathrm{Te}$:n$_\mathrm{HD}$ = 1:3, no DDA, and continuous stirring at 550 rpm). As shown in Fig. ~\ref{fig:MeOH vs time}, the reaction time (y-axis, defined as complete disappearance of TeA) exhibits a clear dependence on the MeOH-to-Te molar ratio (x-axis). The experimental data reveal that the shortest reaction time (\textasciitilde2 h) was achieved at a molar ratio of 80:1. Increasing the ratio to 100:1 maintained the reaction time at \textasciitilde2 h. Further increasing the molar ratio beyond 100:1 resulted in prolonged reaction times. This phenomenon can be attributed to dilution effect. Excessive MeOH reduces the effective concentration of HD thereby decreasing the reaction rate. Hence the optimal MeOH-to-Te molar ratio is (80$\sim$100):1. 

\subsection{Reaction time of Various Diols in MeOH at Room Temperature}
\label{subsubsec:Various dios}
In TeA-diol-MeOH system, TeA can react with various diols at room temperature within several hours even without DDA addition, as shown in Table ~\ref{tab:diols vs time}. In this table, the MeOH-to-Te molar ratio is approximately 40:1, resulting in longer reaction times between TeA and HD compared to the minimum reaction time shown in Fig. ~\ref{fig:MeOH vs time}. Table ~\ref{tab:diols vs time} demonstrates that reaction times gradually increase with elongation of the carbon chain in the diol molecular structure due to enhanced steric hindrance. On the other hand, as the carbon chain lengthens, the lipophilicity of the product increases, thereby enhancing its solubility in organic solvents. Similarly, the addition of DDA to the reaction system significantly reduces the reaction duration.

\begin{table}[!htb]
\begin{center}
\caption{Reaction times of various diols with TeA at room temperature, n$_\mathrm{Te}$:n$_\mathrm{diol}$:n$_\mathrm{MeOH}$=1:3:40.}
\label{tab:diols vs time}
\begin{tabular}{ll}
\hline
Diols         & Reaction time/h
\\
\hline

1,2-propanediol   & \textasciitilde3h
\\1,2-butanediol   & \textasciitilde4h
\\1,2-pentanediol  & \textasciitilde4.8h
\\1,2-hexanediol  & \textasciitilde4.5h
\\1,2-heptanediol  & \textasciitilde5.5h
\\1,2-octanediol  & \textasciitilde6h
\\1,2-nonanediol  & \textasciitilde6.5h\\
\hline
\end{tabular}
\end{center}
\end{table}

\subsection{Mechanistic Study on Methanol-mediated Reaction}
\label{subsubsec:Mechanistic Study}
\begin{figure}[!htb]
\centering
\includegraphics[width=0.9\columnwidth]{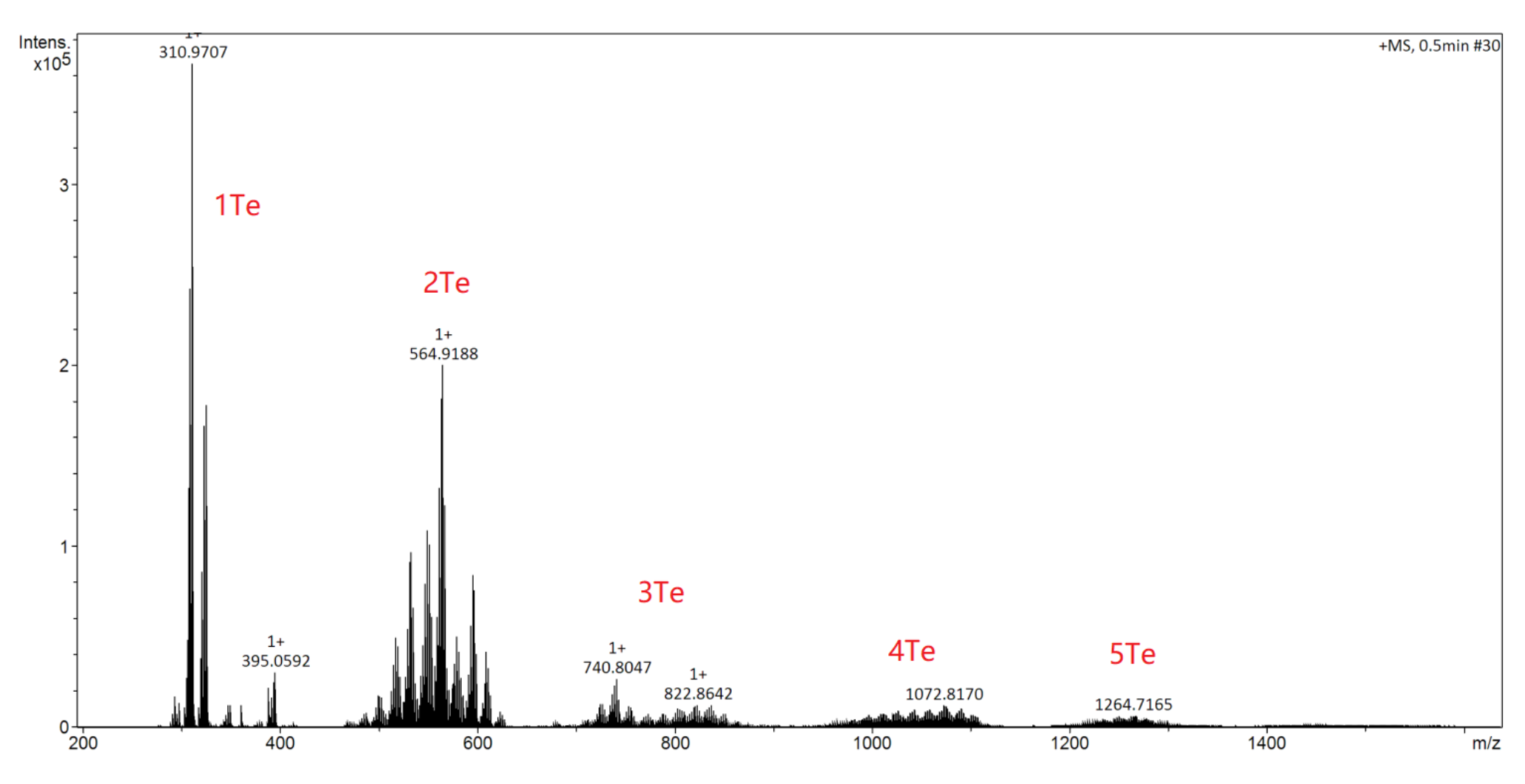}
\caption{Mass spectrum of freshly prepared TeA-MeOH solution.} \label{fig:MS-1}
\end{figure}

\begin{figure}[!htb]
\centering
\includegraphics[width=0.9\columnwidth]{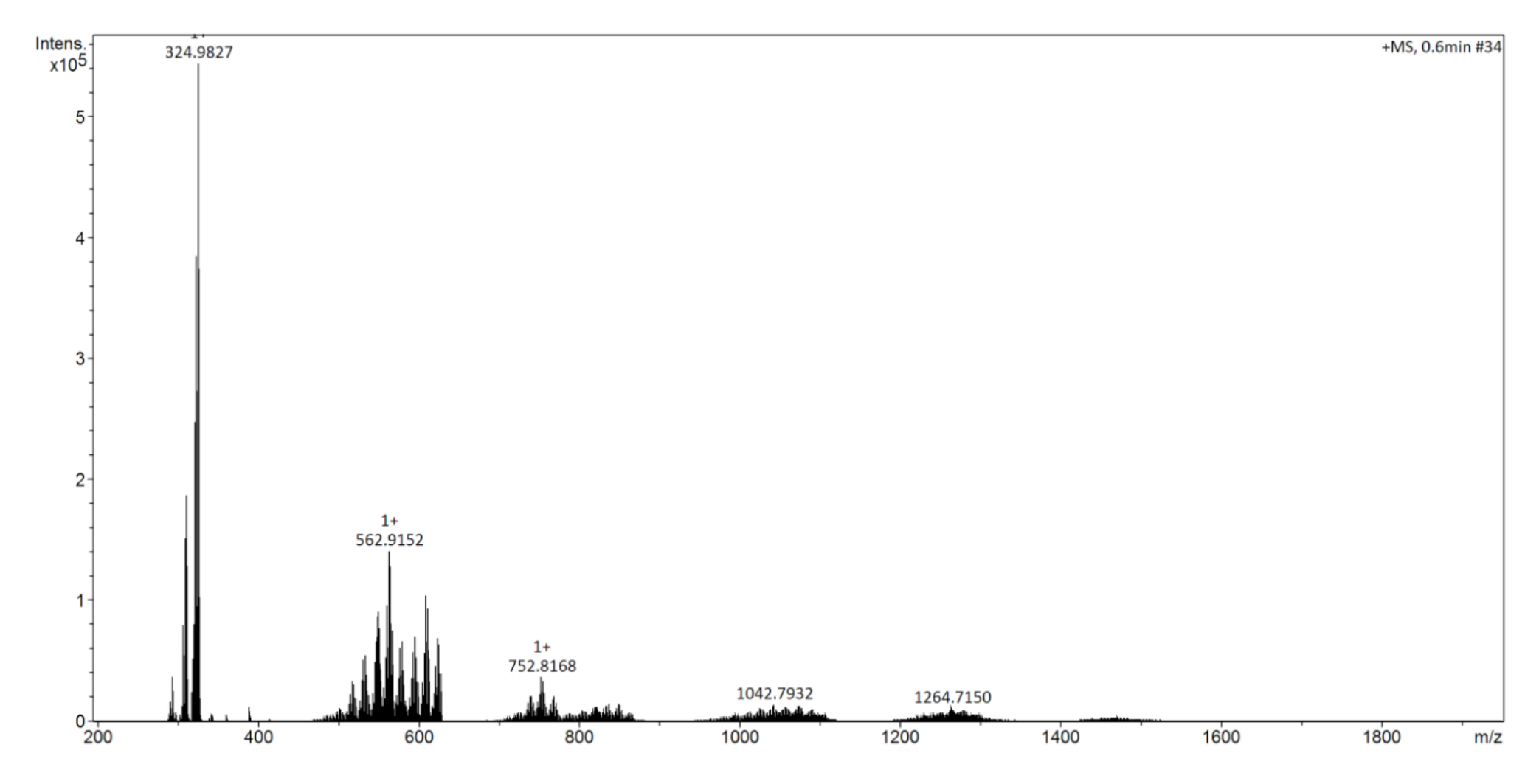}
\caption{Mass spectrum of TeA-MeOH solution after 2.5 years of storage.} \label{fig:MS-2}
\end{figure}

High-resolution time-of-flight mass spectrometry (HR-TOF-MS) was employed to elucidate the reaction mechanism. Initial characterization was performed on a solution of TeA in MeOH. When TeA was stirred in MeOH, the solid gradually appeared to dissolve at room temperature, but this observation resulted from  chemical reactions rather than from simple physical dissolution. HR-TOF-MS analysis of the resulting mixture revealed multiple Te-containing peak clusters in the mass spectrum. Analysis of the Te isotopic abundance ratios in these peak clusters confirmed they correspond to compounds containing 1, 2, 3, 4, and 5 Te atoms in their molecular structures (as shown in Fig. ~\ref{fig:MS-1}). In these poly-tellurium compounds, the Te atoms are not bridged by MeOH molecules. When TeA reacts with HD (which contains two hydroxyl groups), a Te-O-(CH$_2$-CH-(C$_4$H$_9$))-O-Te structure may form, where the HD bridges two Te atoms. In contrast, MeOH possesses only one hydroxyl group and can react with only one TeA molecule, preventing it from bridging two Te atoms. These observations suggest that TeA molecules can undergo dehydration to form Te-O-Te, Te-O-Te-O-Te, and Te-O-Te-O-Te-O-Te structures.
The preliminary structural information of the Te compounds corresponding to the peak clusters was determined by combining Te isotopic abundance ratios and molecular weight data. The relative abundances of the various Te compounds were calculated based on mass spectral peak intensities. For example, The peak cluster spanning m/z 300.97$\sim$310.97 in Fig. ~\ref{fig:MS-1} represents a compound formed by the reaction of one TeA molecule with four MeOH molecules, with elimination of four water molecules, denoted as 1Te(OH)$_6$+4MeOH-4H$_2$O. The four most abundant compounds in the reaction products of TeA with MeOH were 2Te(OH)$_6$+7MeOH-8H$_2$O, 1Te(OH)$_6$+4MeOH-4H$_2$O, 1Te(OH)$_6$+5MeOH-5H$_2$O, and 2Te(OH)$_6$+6MeOH-7H$_2$O, collectively accounting for approximately 65\% of the total abundance. When the Te-MeOH solution was stored at room temperature for 2.5 years and reanalyzed by mass spectrometry, all compound species remained identical, with only minor changes in their relative abundances (Fig. ~\ref{fig:MS-2}). Considering the substantial quantitative errors inherent in mass spectrometry, these results demonstrate that the products derived from the reaction between TeA and MeOH exhibit excellent stability, showing no significant changes after 2.5 years at ambient temperature, indicating high stability.

\begin{figure}[!htb]
\centering
\includegraphics[width=0.9\columnwidth]{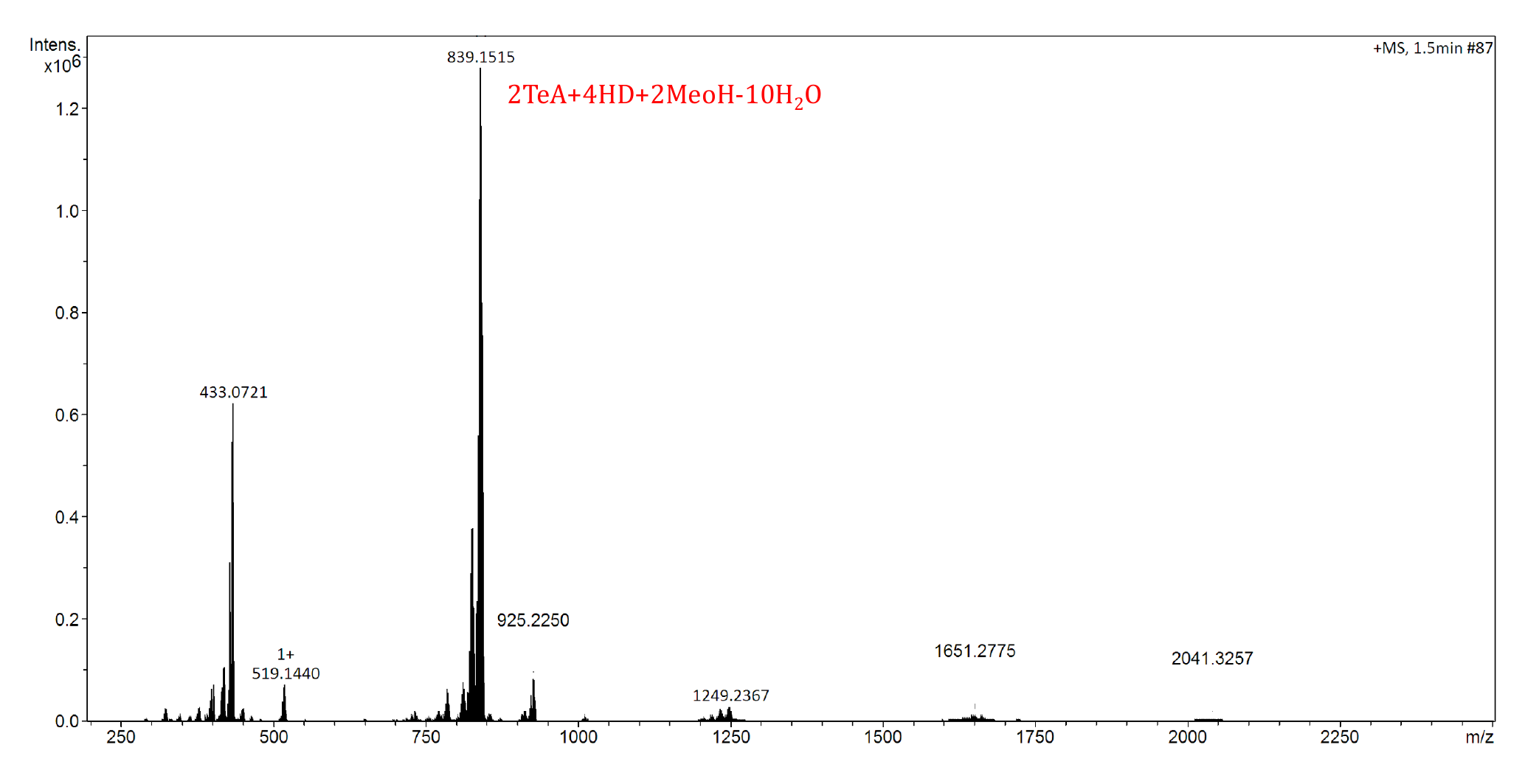}
\caption{Mass spectrum of the TeA–HD reaction product (1:3 molar ratio) in the presence of methanol solvent.} \label{fig:MS-3+1}
\end{figure}

\begin{figure}[!htb]
\centering
\includegraphics[width=0.9\columnwidth]{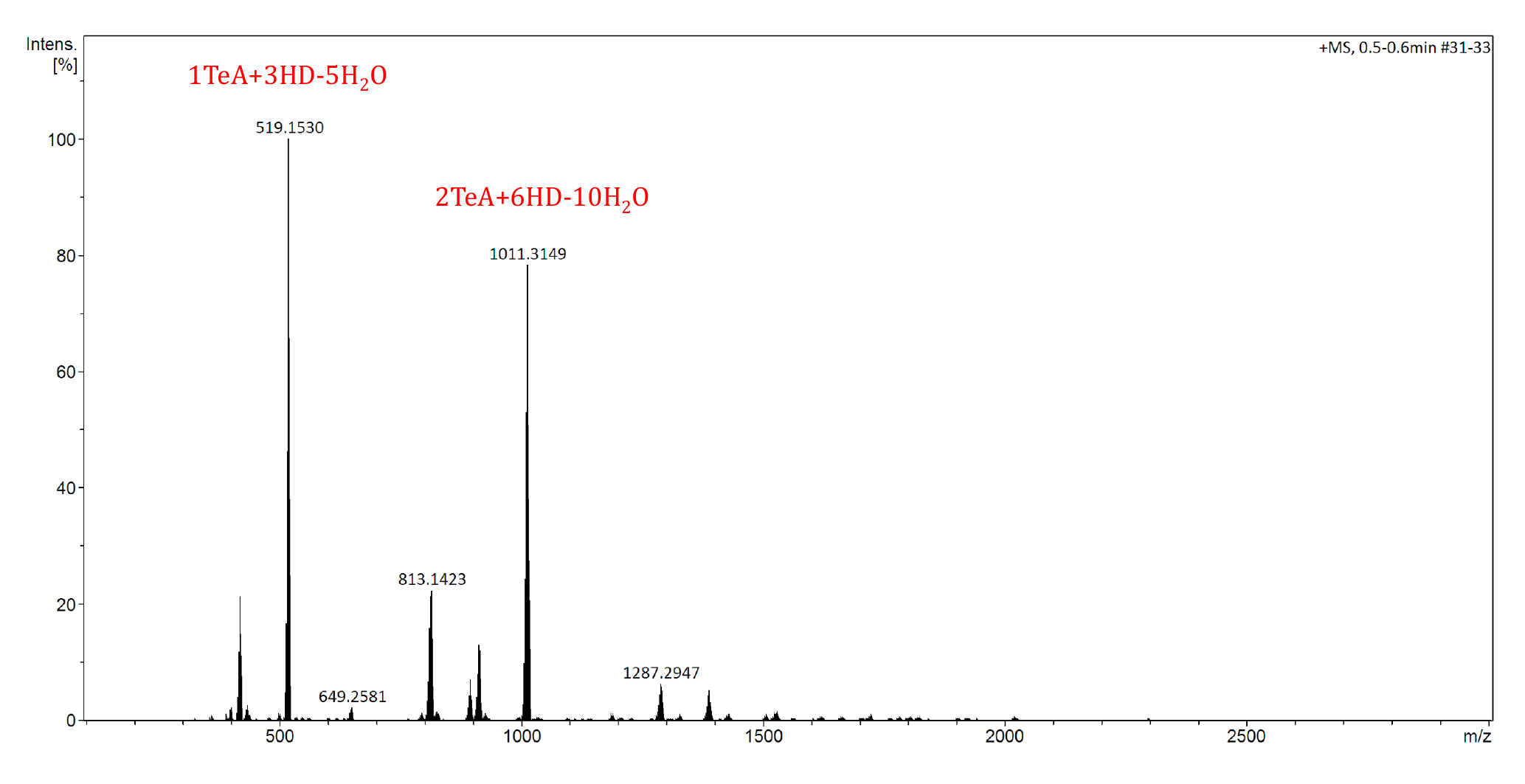}
\caption{Mass spectrum of the TeA–HD reaction product (1:3 molar ratio) after methanol removal.} \label{fig:MS-3+2}
\end{figure}

\begin{figure}[!htb]
\centering
\includegraphics[width=0.9\columnwidth]{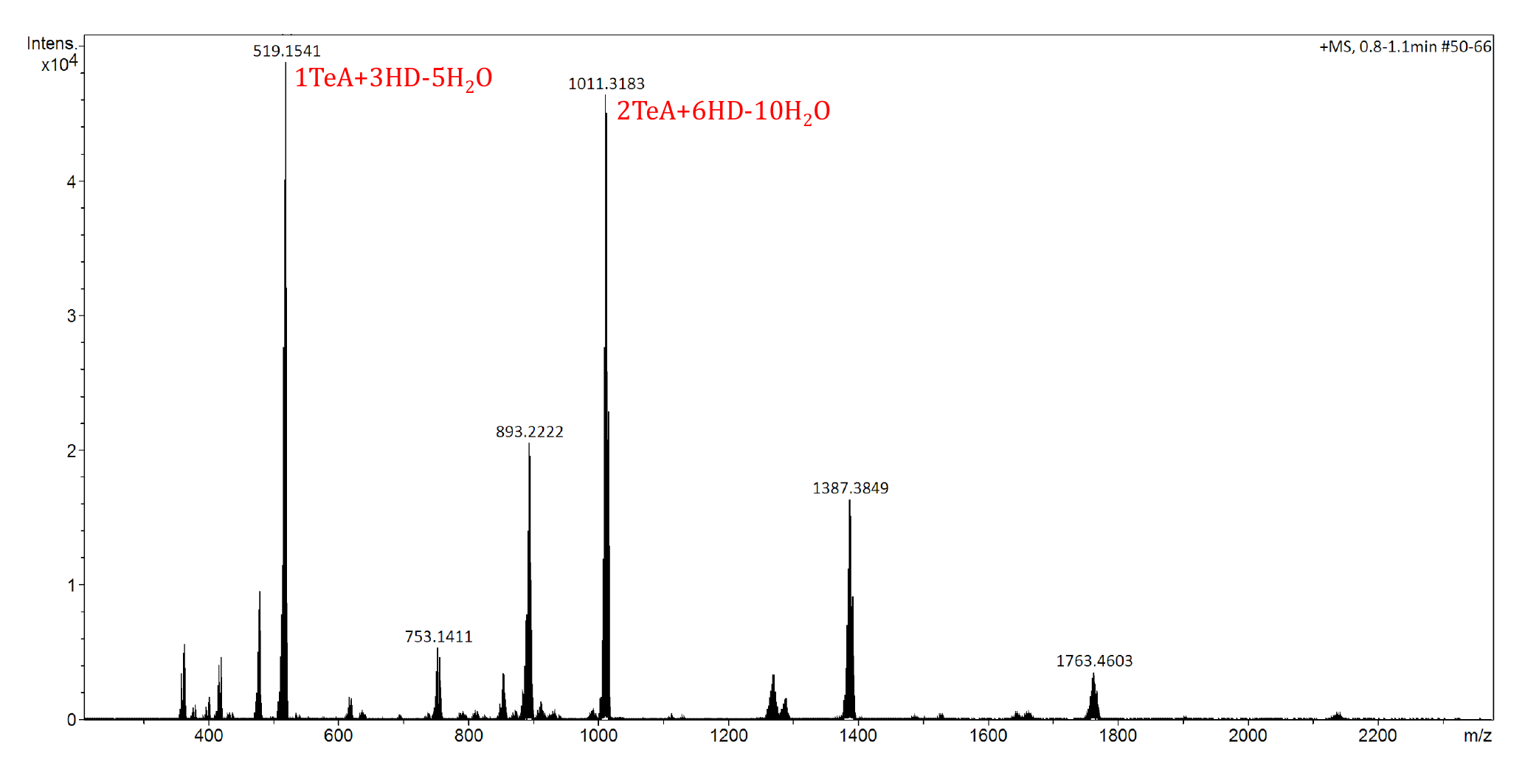}
\caption{Mass spectrum of the TeA-HD reaction product (1:3 molar ratio) synthesized via azeotropic distillation after acetonitrile removal.} \label{fig:MS-3+3}
\end{figure}

\begin{figure}[!htb]
\centering
\includegraphics[width=0.9\columnwidth]{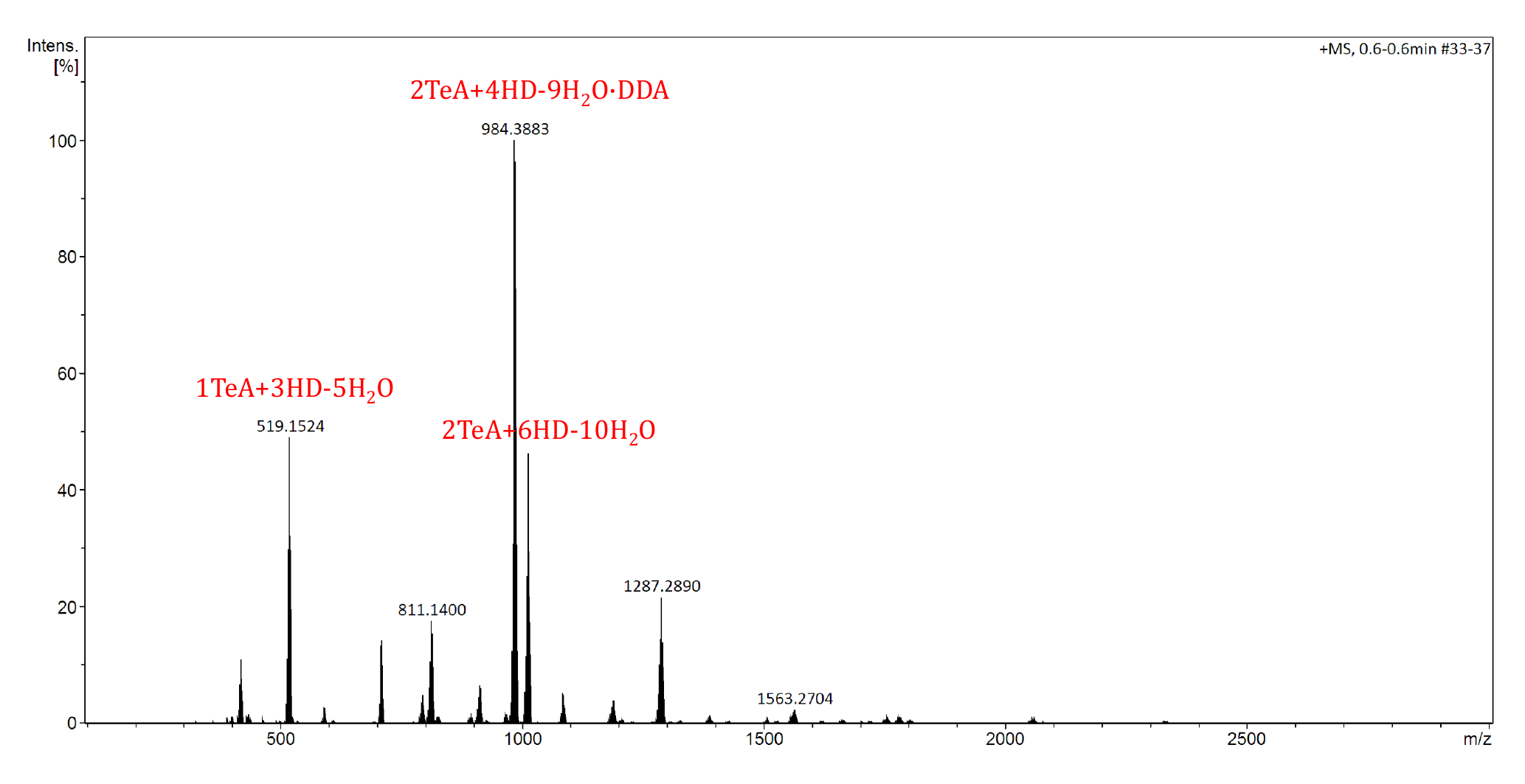}
\caption{Mass spectrum of the reaction product of TeA, HD, and DDA at a molar ratio of 1:3:0.2 (after methanol removal)} \label{fig:MS-3+4}
\end{figure}

Subsequently, the Te-diol compounds were synthesized and characterized by mass spectrometry to identify key intermediates and products. The reaction was performed at room temperature using MeOH as solvent with a molar ratio of 1:3 TeA to HD. Mass spectrometric analysis of the crude reaction mixture (without MeOH removal) revealed a dominant 2Te-containing compound accounting for \textgreater 50\% relative abundance. Based on molecular weight data, this species was tentatively assigned as 2TeA+4HD+2MeOH-10H$_2$O as shown in Fig. ~\ref{fig:MS-3+1}. Notably, no compounds derived solely from TeA and MeOH were detected in the products.
A comparative synthesis was conducted under identical conditions, with the sole modification being post-reaction removal of MeOH via vacuum distillation (25$\pm$5°C) until constant product weight was achieved. The Te content of the product obtained in this manner was consistent with that of the product obtained after removing acetonitrile via azeotropic distillation. This demonstrates that the methanol solvent has been substantially removed. Following the achievement of constant weight, mass spectrometric analysis revealed that the principal products consisted exclusively of Te-HD compounds, with neither 2TeA+4HD+2MeOH-10H$_2$O nor Te-MeOH derivatives detected, as shown in Fig.~\ref{fig:MS-3+2}. The two predominant species exhibiting the highest relative content are identified as 2TeA+6HD-10H$_2$O and 1TeA+3HD-5H$_2$O; this finding aligns with the two major products observed in the azeotropic distillation process (as shown in Fig.~\ref{fig:MS-3+3}).  
Therefore, although both MeOH and HD can react with TeA, distillation removal of MeOH after reaction completion effectively eliminates its influence on product composition, yielding exclusively Te-HD compounds. This demonstrates that Te-MeOH compounds exhibit lower stability compared to Te-diol compounds, which is chemically reasonable since diols, possessing two hydroxyl groups, can form stable five-membered cyclic structures with TeA.
Based on the mass spectrometry results, the reaction mechanism is proposed as follows: In the TeA/HD/MeOH system, MeOH preferentially reacts with TeA due to its large excess and minimal steric hindrance, generating a series of reactive and unstable intermediates. In the absence of HD, these intermediates would further react with MeOH to form relatively stable Te-MeOH compounds (see Fig. ~\ref{fig:MS-1}). When HD is present, the intermediates react with it to produce more stable Te-HD-MeOH and Te-HD compounds. When ethanol is used as solvent, it also reacts with TeA and exhibits acceleration effects, though the reaction time increases significantly due to ethanol's greater steric hindrance compared to MeOH. For 1-propanol, the acceleration effect is further reduced because of even larger steric hindrance. During vacuum distillation after reaction completion, the less stable Te-HD-MeOH compounds react further with free HD remaining in the system to form Te-diol compounds. In contrast, solvents like acetonitrile and chloroform do not react with or dissolve TeA, leading to significantly slower reaction rates in this heterogeneous reaction. Therefore, MeOH exhibits certain catalyst-like characteristics: it participates in the reaction, accelerates the reaction rate, and does not alter the structure of the final product. However, MeOH is not a classical catalyst. Its role somewhat resembles that of a phase-transfer catalyst, though it is not entirely identical. Given this, the room-temperature reaction between TeA and HD in MeOH is termed a "methanol-mediated reaction. "

When DDA was added to the reaction system (n$_\mathrm{Te}$:n$_\mathrm{HD}$:n$_\mathrm{DDA}$ = 1:3:0.2), the MeOH was removed by vacuum distillation after reaction completion, followed by mass spectrometric characterization of the products. Mass spectrometric analysis revealed that the products consisted exclusively of Te-HD compounds, with a possible composition of 2TeA+4HD-9H$_2$O (in the form of DDA complex) being the most abundant species, followed by two other species with relatively high content: 2TeA+6HD-10H$_2$O and 1TeA+3HD-5H$_2$O, as shown in Fig. ~\ref{fig:MS-3+4}. Results from multiple syntheses and characterizations indicate that as the molar ratio of DDA to Te increases stepwise (0.05, 0.1, and 0.2), the relative content of 2TeA+4HD-9H$_2$O (in the form of DDA complex) increases accordingly. This indirectly verifies that DDA is incorporated into the composition of this 2Te compound. This result was consistent with the mass spectrometric characterization of products synthesized via azeotropic distillation using acetonitrile as the solvent under identical reactant ratios, except that the relative contents of the three major products were somewhat different.

\section{Optimization of DDA as stabilizer}
\label{sec:Alkylamine}

In the reaction system of TeA and diols, the SNO+ collaboration pioneered the use of DDA as a stabilisation/solubilisation agent for Te-BD~\cite{SNO+:2023}. For the room-temperature synthesis method (as well as the azeotropic distillation method), the TeA and HD system was selected. In the TeA-HD reaction system, where the product Te-HD is miscible with LAB, the inclusion of DDA accelerates the reaction but primarily acts as a stabilizer.

\begin{figure}[!htb]
\centering
\includegraphics[width=0.45\columnwidth]{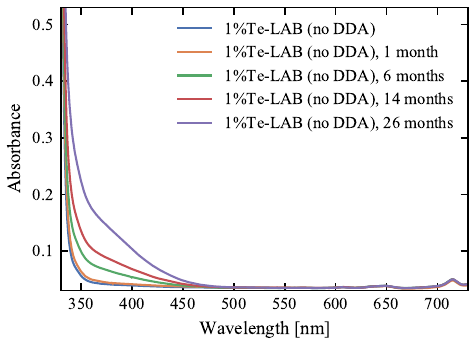}
\includegraphics[width=0.45\columnwidth]{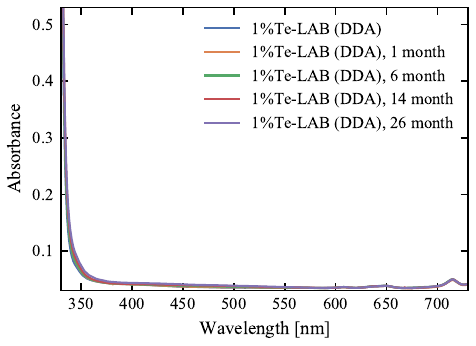}
\caption{Absorption spectra of 1\% Te-LAB solutions: (a) prepared without DDA addition during synthesis (left panel); (b) prepared with DDA addition during synthesis (right panel).} \label{fig:Te-LAB-DDA}
\end{figure}

The Te-HD compounds were dissolved in the LS solvent LAB to prepare solutions with specific Te concentrations. 
The long-term stability was evaluated through periodic absorption spectral measurements of Te-LAB solutions, conducted using a PE Lambda 850+ UV-Vis spectrophotometer (PerkinElmer, USA) equipped with a 10 cm quartz cell with the absorbance sensitivity of 10$^{-4}$. 
Multiple synthesis trials consistently demonstrated that Te-LAB samples prepared via room-temperature synthesis using MeOH as solvent without DDA addition exhibited poor long-term stability. Fig.~\ref{fig:Te-LAB-DDA}-a shows the absorption spectra of a 1\% Te-LAB solution prepared from DDA-free synthesis products, where the absorbance below 470 nm increased significantly over time, indicating poor sample stability. In contrast, the addition of DDA during synthesis substantially improved the long-term stability of Te-LAB samples (Fig.~\ref{fig:Te-LAB-DDA}-b). These two syntheses were conducted simultaneously under identical conditions - same reactants (TeA, HD and MeOH), same stoichiometric ratios (n$_\mathrm{Te}$:n$_\mathrm{HD}$:n$_\mathrm{MeOH}$ = 1:3:42), and same stirring speed - with the sole difference being the addition of DDA (n$_\mathrm{Te}$:n$_\mathrm{DDA}$ = 1:0.2) in ~\ref{fig:Te-LAB-DDA}-b but not in ~\ref{fig:Te-LAB-DDA}-a. The exact cause of the increase in sample absorbance in the absence of DDA remains undetermined due to insufficient experimental data. To address this, we intend to employ a multi-analytical approach for further investigation. For example, mass spectrometry will be utilized to analyze the structural and quantitative changes of Te compounds before and after the absorbance increase. Another planned approach involves applying sample pretreatment coupled with GC-MS will be applied to detect any potential trace light-absorbing impurities generated via chemical reactions within the sample. Furthermore, long-term monitoring of the light yield stability will be conducted to verify the homogeneity of Te compounds in the sample.

\begin{figure}[!htb]
\centering
\includegraphics[width=0.45\columnwidth]{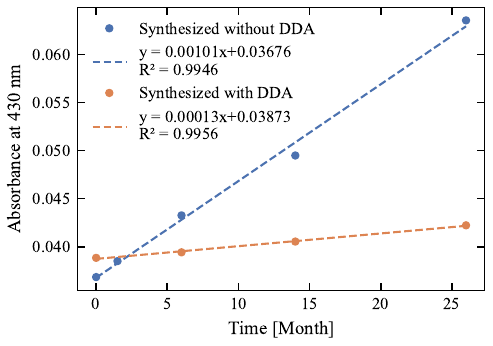}
\caption{Temporal evolution of absorbance at 430 nm for 1\% Te-LAB samples.} \label{fig:430-W-O-DDA-Syn}
\end{figure}

In addition to absorption spectra, the temporal evolution of absorbance at 430 nm (the sensitive region for PMT detection) was monitored, see Fig. ~\ref{fig:430-W-O-DDA-Syn}. For 1\% Te-LAB solutions, the monthly absorbance increase was $(10.1\pm0.1)\times10^{-4}$ without DDA, while With the addition of DDA, it was reduced by nearly an order of magnitude to $(1.3\pm0.1)\times10^{-4}$. The error is induced from the stability of the equipment, characterized by the standard deviation of the measured absorbance at 430 nm from a LAB control sample through the courses of months. Both spectral and absorbance monitoring results clearly demonstrate the stabilizing effect of DDA on the optical properties. Although the monthly absorbance increase of the DDA-added samples does not yet fully meet the requirements of the JUNO $0\nu\beta\beta$ experiment, it should be noted that these do not represent the optimized samples. Sec.~\ref{subsubsec:AL} presents samples with significantly higher transparency, whose long-term stability will be reported in the future.

\begin{figure}[!htb]
\centering
\includegraphics[width=0.45\columnwidth]{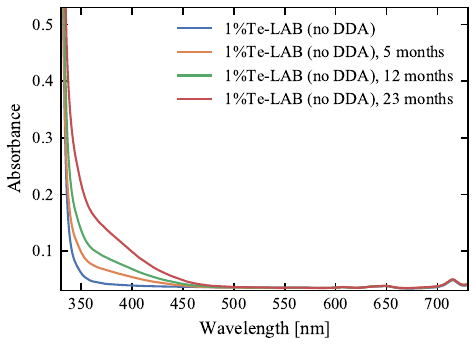}
\includegraphics[width=0.45\columnwidth]{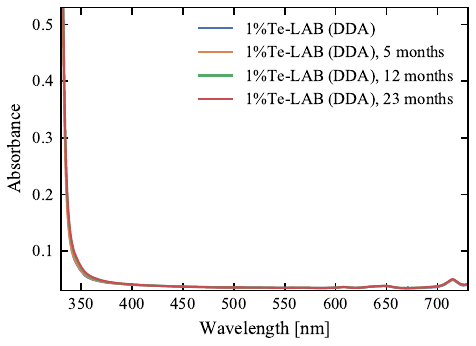}
\caption{ Absorption spectra of 1\% Te-LAB solutions prepared under different conditions: (a) without DDA (left) and (b) with DDA introduced during preparation (right).} \label{fig:Te-LAB-DDA_preparation}
\end{figure}

\begin{figure}[!htb]
\centering
\includegraphics[width=0.45\columnwidth]{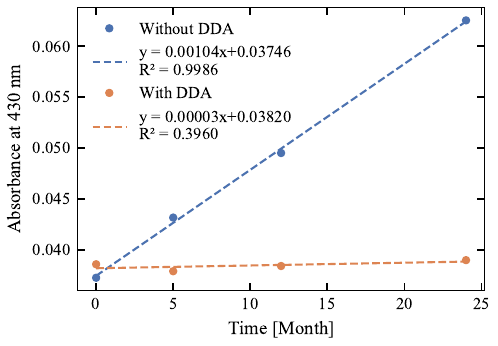}
\caption{Temporal evolution of absorbance at 430 nm for 1\% Te-LAB samples prepared with/without DDA} \label{fig:430-W-O-DDA-prep}
\end{figure}

The approach of adding DDA during synthesis to enhance product stability offers the advantage of accelerated reaction rates but suffers from partial DDA loss during solvent removal. Both vacuum distillation and nitrogen purging for solvent elimination inevitably remove some DDA along with MeOH, as confirmed by GC-MS analysis. This variation in DDA concentration in the final product adversely affects batch-to-batch consistency and reproducibility. An alternative DDA incorporation method that was also mentioned in Ref.~\cite{SNO+:2023} was therefore applied: DDA was excluded during synthesis but quantitatively added when preparing Te-LAB solutions using the Te-HD compounds. Experimental results demonstrate that this approach equally improves Te-LAB stability ( Fig.~\ref{fig:Te-LAB-DDA_preparation}). The 1\% Te-LAB solutions measured in Fig.~\ref{fig:Te-LAB-DDA_preparation}-a and Fig.~\ref{fig:Te-LAB-DDA_preparation}-b originated from the same DDA-free synthesis batch, differing only in that the latter contained DDA (n$_\mathrm{Te}$:n$_\mathrm{DDA}$=1:0.2) added during solution preparation. Two-year monitoring revealed significantly enhanced stability for the DDA-containing solution. Fig. ~\ref{fig:430-W-O-DDA-prep} shows the monthly absorbance increases at 430 nm decreased from $(10.4\pm0.1)\times10^{-4}$ (without DDA) to $(0.3\pm0.1)\times10^{-4}$ (with DDA), consistent with results obtained from synthesis-stage DDA addition.

\begin{figure}[!htb]
\centering
\includegraphics[width=0.45\columnwidth]{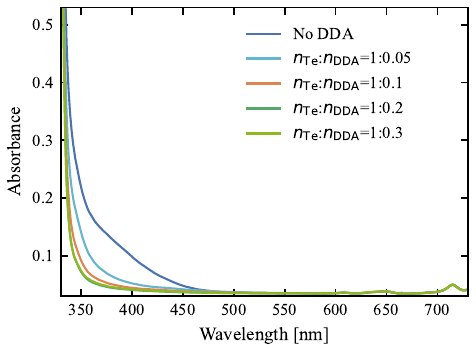}
\caption{Absorption spectra of 1\% Te-LAB with different DDA concentrations stored for 23 months at ambient temperature.} 
\label{fig:DDA-optimization}
\end{figure}

Optimization of DDA dosage was performed using 1\% Te-LAB solutions. Fig.~\ref{fig:DDA-optimization} presents the results of samples aged at room temperature for 23 months, where the optimal effect was achieved when the molar ratio of DDA to Te ranged between 0.2 and 0.3. Although spectral data for higher DDA dosages were initially lacking, the strict requirements of the JUNO Central Detector (CD) for radioactive background favor the use of minimal DDA; thus, from this perspective, experiments with higher DDA dosages were not strictly necessary. Nevertheless, experiments at room temperature with higher DDA and Te concentrations are currently underway, but the required experimental time is lengthy, and the results will be reported in subsequent studies.

\section{Performance Characterization of Room-Temperature Synthesis Products}
\label{subsec:Characterization}

\subsection{Transparency characterized by UV-Vis spectrophotometer}
\label{subsec:UV-Vis spectra}
\subsubsection{UV-Vis absorption spectrum}
\label{subsec:spectra}

\begin{figure}[!htb]
\centering
\includegraphics[width=0.45\columnwidth]{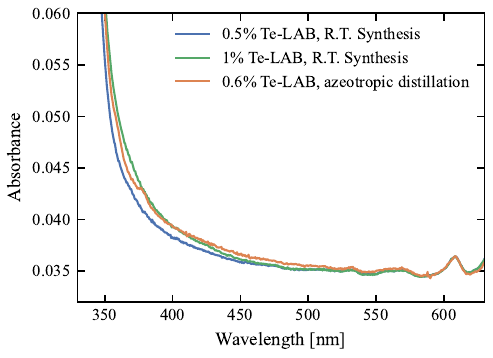}
\caption{Comparison of spectra of Te-LAB samples prepared by room-temperature synthesis and azeotropic distillation method.} \label{fig:comparison}
\end{figure}

To clearly demonstrate the effect of Te-HD compounds on transparency, measurements were conducted without adding any fluors, using only highly purified LAB to dissolve the Te-HD compounds for UV-Vis absorption spectroscopy. All LAB solvents were purified by Al$_2$O$_3$, exhibiting excellent absorption spectra with attenuation lengths of 24$\sim$25 m at 430 nm.
No significant difference in spectral performance was observed between Te-LAB samples prepared from room-temperature synthesized compounds and products from the azeotropic distillation approach. In Fig.~\ref{fig:comparison}, the green line represents the spectrum of 0.6\% Te-LAB prepared via azeotropic distillation, while the blue and orange curves correspond to spectra of Te-LAB samples (with proportional DDA addition) prepared by room-temperature synthesis at Te concentrations of 0.5\% and 1\%, respectively. The room-temperature synthesized samples demonstrate spectral characteristics comparable to those synthesized via the azeotropic distillation method. 

\subsubsection{Absorbance at 430nm}
\label{sec:Abs-430}

\begin{figure}[!htb]
\centering
\includegraphics[width=0.45\columnwidth]{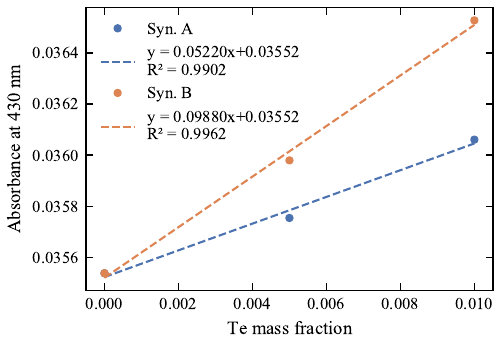}
\includegraphics[width=0.45\columnwidth]{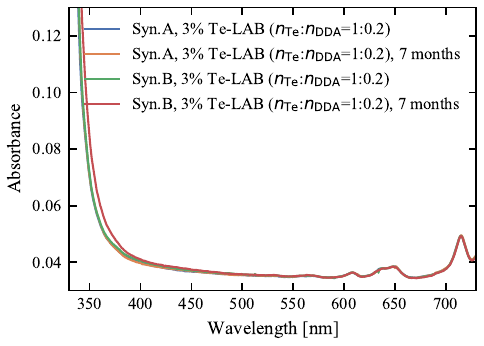}
\caption{(a) Synthesis A\&B: Te concentration-430 nm absorbance correlation (left panel); (b) 3\% Te-LAB spectra with/without DDA from Synthesis A\&B (right panel).} \label{fig:Synthesis A&B}
\end{figure}

At 430 nm - the sensitive region for PMT detection - it is essential to investigate the influence of Te concentration on absorbance. Solutions with varying Te concentrations in LAB were prepared and their absorbance at 430 nm measured. Plotting absorbance versus Te mass percentage yields a linear relationship, where the slope represents the rate of absorbance change with Te content. A greater slope indicates poorer transparency of the Te-LAB sample.
Two room-temperature syntheses (A and B) were performed using MeOH as solvent with n$_\mathrm{Te}$:n$_\mathrm{HD}$=1:3, both without DDA addition. While employing identical reactants, the syntheses differed in reaction time. At the point when synthesis A was completed, a trace amount of solid TeA particles remained in synthesis B. This is understandable given that this was a small-scale exploratory synthesis; although the solid TeA was ground using a mortar and pestle, the particle size was not precisely controlled. Consequently, the reaction time for synthesis B had to be extended. However, as the reaction was nearing completion, the concentration of HD in the system had significantly decreased. Combined with the small amount of remaining TeA particles and the resulting minimal contact area, this ultimately led to the reaction time for synthesis B being twice that of synthesis A. The products were diluted with LAB to prepare 0.5\% and 1\% Te-LAB solutions for absorption spectroscopy measurements. Given the high transparency of all samples, with absorption spectra approaching the instrument detection limit between 400-600 nm, discernible differences between spectra of equal Te concentration could not be conclusively determined. However, examination of 430 nm absorbance revealed that for each 1\% increase in Te concentration, synthesis B showed an absorbance increase of $(9.9\pm0.1)\times10^{-4}$, while synthesis A exhibited an increase of only $(5.2\pm0.1)\times10^{-4}$ (as shown in Fig.~\ref{fig:Synthesis A&B}-a). This preliminary data suggested better transparency for synthesis A products. For verification, both synthesis products were prepared as 3\% Te-LAB solutions with DDA addition (n$_\mathrm{Te}$:n$_\mathrm{DDA}$=1:0.2). Initial absorption spectra showed no significant differences, but after 7 months, synthesis B demonstrated markedly increased absorbance compared to synthesis A (Fig.~\ref{fig:Synthesis A&B}-b), confirming the superior transparency of synthesis A products. It is speculated that the prolonged reaction time in synthesis B may have led to the formation of Te-HD products with a higher degree of polymerization, or a higher abundance of species that contribute to optical absorbance. This phenomenon is also prevalent in azeotropic distillation methods, where an extended reaction time typically results in poorer spectral quality of the prepared Te-LAB. In any case, these results validate using the rate of 430 nm absorbance change with Te concentration as a reliable indicator of Te solution transparency.

\begin{figure}[!htb]
\centering
\includegraphics[width=0.45\columnwidth]{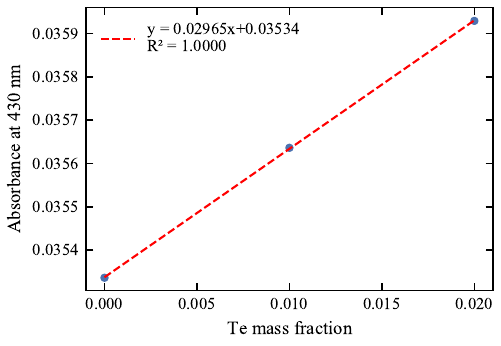}
\caption{Correlation between Te content and absorbance at 430 nm for Te-LAB solutions prepared from the current optimal room-temperature synthesis product.} \label{fig:Te-abs}
\end{figure}

The optimal synthesis product obtained previously was derived from hundred-kilogram-scale Te-LS production. As the full-scale hundred-kilogram production is not yet complete, this paper presents results only from the first batch of Te-LAB samples: the rate of increase in absorbance at 430 nm with respect to tellurium concentration is $(3.0\pm0.1)\times10^{-4}$ /1\% (see Fig.~\ref{fig:Te-abs}). In the other batches, the rate of increase in absorbance at 430 nm with respect to Te concentration remains on the order of $10^{-4}$ per 1\% Te, with some samples showing values as low as $(2.0\pm0.1)\times10^{-4}$/1\% Te. Further results will be presented in future publications detailing the full hundred-kilogram-scale Te-LS production.

\subsection{Long-term stability monitored by absorption spectra}
\label{sec:Long-term stability}

\begin{figure}[!htb]
\centering
\includegraphics[width=0.45\columnwidth]{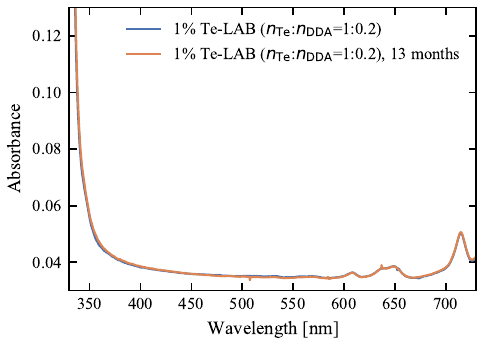}
\includegraphics[width=0.45\columnwidth]{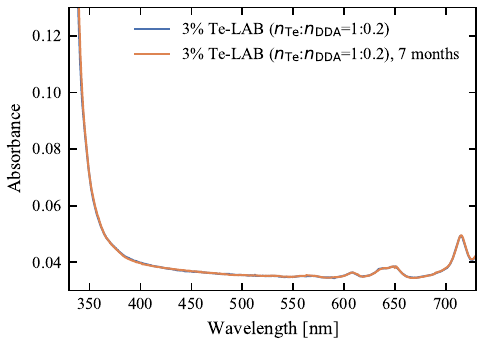}
\caption{Long-term stability of 1\%Te and 3\%Te-LAB, monitored by absorption spectra.} \label{fig:stability}
\end{figure}

Several syntheses were performed at room temperature using MeOH as solvent without DDA addition (n$_\mathrm{TeA}$:n$_\mathrm{HD}$ = 1:3). The products were subsequently prepared as Te-LAB solutions with DDA addition (n$_\mathrm{Te}$:n$_\mathrm{DDA}$ = 1:0.2). To prevent sample degradation from measurement-induced contamination, each Te-LAB sample was aliquoted into multiple vials, with one vial used for each spectral measurement before disposal. Both 1\% and 3\% Te-LAB solutions have maintained excellent spectral stability to date (see Fig.~\ref{fig:stability}).

\begin{figure}[!htb]
\centering
\includegraphics[width=0.45\columnwidth]{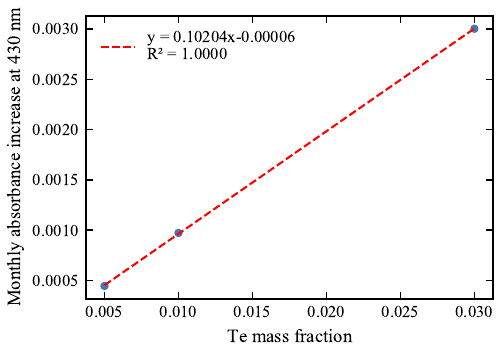}
\caption{Relationship between the monthly increase in absorbance at 430 nm and Te content in Te-LAB solution synthesized without DDA.} \label{fig:Abs-mths vs Te}
\end{figure}

One DDA-free room-temperature synthesis product was diluted with LAB to prepare three samples with Te concentrations of 0.5\%, 1\%, and 3\%; no DDA was involved during the whole procedure. These samples were monitored via absorption spectroscopy for two years (with repeated measurements of the same samples), yielding the monthly absorbance increase at 430 nm for each sample. Plotting these monthly increases against Te concentration produced a linear relationship (see Fig.~\ref{fig:Abs-mths vs Te}). The results demonstrate that for DDA-free solutions with poorer stability, each 1\% increase in Te concentration corresponds to a $(10.2\pm0.1)\times10^{-4}$ increase in the monthly absorbance increment at 430 nm. This linear relationship enables prediction of the monthly absorbance increase at 430 nm for DDA-free samples across different Te concentrations.

With DDA addition, the monthly increase in 430 nm absorbance remains relatively low across Te-LAB samples of different concentrations. Since the monitoring period currently spans only seven months (Fig.~\ref{fig:stability}), the relationship between Te content and monthly absorbance increase at 430 nm cannot yet be established.

\begin{figure}[!htb]
\centering
\includegraphics[width=0.45\columnwidth]{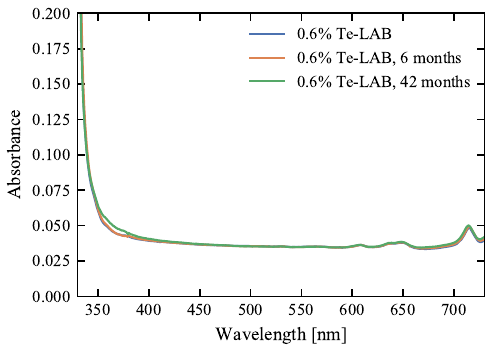}
\includegraphics[width=0.45\columnwidth]{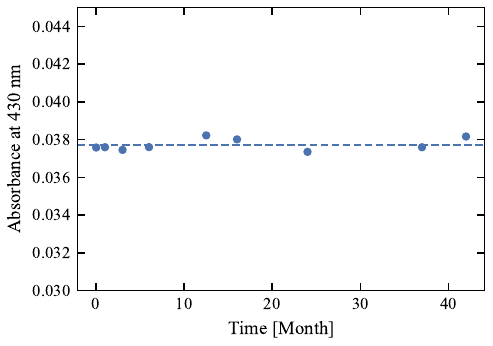}
\caption{Absorption spectra (left panel) and time-dependent absorbance at 430 nm (right panel) of 0.6\% Te-LAB synthesized by azeotropic distillation approach.} \label{fig:AD synthesis}
\end{figure}

Our previously reported azeotropic distillation approach ~\cite{IHEP:2023} shows updated long-term stability data for Te-LAB solutions. After 3.5 years, a slight absorbance increase was observed in the 350-385 nm range, potentially due to repeated sample opening for measurements causing gradual aging. However, the absorbance at 430 nm remained essentially stable throughout the three-year period (see Fig.~\ref{fig:AD synthesis}).

\subsection{Attenuation length}
\label{subsubsec:AL}

To characterize the transparency of Te-LS, we employ two methods: absorption spectroscopy and attenuation length (A.L.) measurement. Each method has its advantages and limitations, and both are indispensable for large-scale detectors like JUNO. Absorption spectroscopy is rapid and requires only a small sample volume, making it suitable for the preliminary research phase to screen samples quickly. However, for highly transparent liquids, measurements with a 10 cm optical path are insufficiently sensitive; therefore, we do not intend to calculate the A.L. using 10 cm absorbance data. Once samples with favorable absorption spectra are identified, the synthesis scale is increased to obtain a sufficient quantity for A.L. measurements.

Recently, we have been conducting hundred-kilogram-scale Te-LS production. Although the full production run is not yet complete, we have already obtained hundreds of Te-loaded samples. Consequently, we measured these samples using the 1-meter optical path attenuation length measurement system (at a wavelength of 430$\pm$5 nm) employed in JUNO LS production ~\cite{ Yin:2020elg, Zhu:2022jwc}. The measurement results for selected samples are as follows: the attenuation length for 1\% Te-LAB is 20.1 $\pm$1.1 m; for 2\% Te-LAB, it is 18.6 $\pm$0.9 m; and for 3\% Te-LAB, it is 17.8 $\pm$0.9 m. These results indicate that the synthesis conditions of the 100 kg-scale production are well-controlled, yielding samples with high transparency.

To date, no other direct measurements of A.L. data for Te-loaded samples have been reported. However, spectral data from the SNO+ collaboration suggests that an extinction length exceeding 10 m at 430 nm is achievable with 10\% tellurium loading ~\cite{ SNO+:2023}. As the hundred-kilogram-scale Te-LS production continues to be optimized, characterization data for samples with varying Te concentrations will become available and will be reported in subsequent publications.

\subsection{Relative light yield}
\label{subsubsec:LY}
The light yield of the Te-loaded sample is characterized in terms of relative light yield with regard to an unloaded LS sample (an LS sample with identical composition except for the absence of Te). This was measured by a setup developed specifically for batch characterization of Te-LS samples \cite{Han:2025}. It mainly utilizes the conversion electrons (around 1 MeV) from a $^{207}$Bi source. The electrons deposit energy to Te-LS in a sample bottle with a volume of around 20 mL. 

\begin{figure}[!htb]
\centering
\includegraphics[width=0.45\columnwidth]{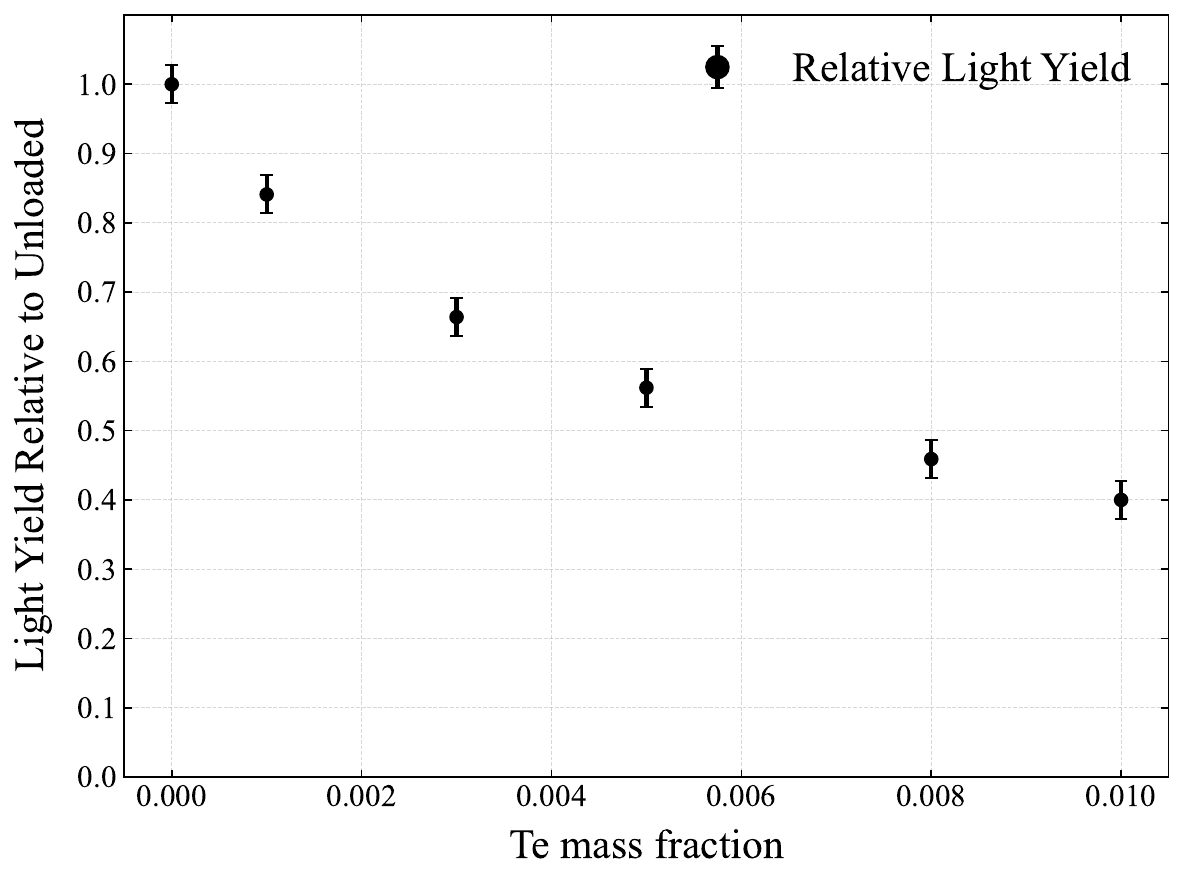}
\caption{Relative light yields as a function of Te loading for LAB with 2.5 g/L PPO and 3mg/L bis-MSB. } \label{fig:LY}
\end{figure}

The relationship between the light yield of Te-LS and Te content is shown in Fig.~\ref{fig:LY}. The 0.5\% loaded Te-LS sample (Te-LAB with 2.5 g/L 2,5-Diphenyloxazole (PPO) and 3 mg/L 1,4-Bis(2-methylstyryl)benzene (bis-MSB)) gives a light yield of 56\%±2\%, in agreement with the sample prepared by the azeotropic distillation method in our previous work \cite{IHEP:2023}. Additionally, the light yield profile of Te-LS obtained by the type I loading method reported in Ref.\cite{SNO+:2023} shows very similar results to Fig.~\ref{fig:LY}. This similarity is primarily attributed to the structural and compositional resemblance between the synthetic products in this work and those obtained by the type I loading method. In our synthetic products, the two compounds with the highest relative abundance are 2TeA+6HD-10H$_2$O and 1TeA+3HD-5H$_2$O. The type I loading method products show the two most abundant compounds as 2Te+5BD-10H$_2$O and 1TeA+3BD-5H$_2$O.The 2TeA+6HD-10H$_2$O compound in this work has a corresponding structure 2TeA+6BD-10H$_2$O in the literature, while the 2Te+5BD-10H$_2$O compound has a corresponding structure 2Te+5HD-10H$_2$O in our synthetic products. In summary, the structures of our synthetic products and those from the type I loading method show significant overlap, both containing considerable amounts of oligomeric compounds, which consequently exhibit strong fluorescence quenching effects on the liquid scintillator. The type II loading method in Ref.\cite{SNO+:2023}, by avoiding heat application during synthesis and post-processing to prevent oligomerization and selecting specific reactant ratios (a molar ratio of 0.5:1:2 for DDA:TeA:BD), obtained products composed almost entirely of 1Te compounds (1TeA+2BD-4H$_2$O), thereby reducing fluorescence quenching in the liquid scintillator and achieving higher light yield: at a Te loading of 1.8\%, the light yield is approximately half that of the initial scintillator. This research finding provides a very valuable approach for improving the light yield of Te-LS.

\section{Conclusions and discussions}
\label{sec:summary}
An energy efficient synthesis method for Te-diol compounds (with a primary focus on Te-HD) was developed in this study.  Compared with our previous azeotropic distillation method, this method reduces energy consumption and mitigates safety hazards while maintaining batch reproducibility and scalability. The approach involves direct mixing of solid TeA, diols, and MeOH under ambient conditions.  In this methanol-mediated reaction system, MeOH actively participates in the reaction and accelerates the reaction rate, thereby enabling this heterogeneous reaction to be completed within an acceptable timeframe. Under these mild conditions (n$_\mathrm{Te}$:n$_\mathrm{HD}$:n$_\mathrm{MeOH}$=1:3:80, TeA=0.5 kg), the reaction completes within 4 hours at room temperature with stirring. Notably, adding a small amount of DDA (0.2 molar equivalents relative to Te) further reduces the reaction time to 2 hours. The process concludes with mild-temperature vacuum distillation to remove MeOH, yielding Te-diol compounds that are directly applicable for Te-LS preparation.

Experimental investigations into the MeOH-mediated mechanism suggest that MeOH preferentially reacts with TeA to form intermediates, which subsequently react with HD to produce Te-diol-MeOH compounds. During post-reaction MeOH removal, the Te-diol-MeOH compounds convert into more stable Te-diol compounds. These observations demonstrate MeOH's dual role: serving as a solvent while exerting a catalytic effect. Subsequent optimization established an optimal MeOH dosage range of n$_\mathrm{MeOH}$:n$_\mathrm{Te}$ = (80$\sim$100):1.

DDA plays a crucial role as a stabilization agent in the Te-LAB solution prepared from Te-HD. 
Without DDA, the 1\% Te-LAB solution deteriorates with a monthly absorbance increase on the order of $(10\pm0.1)\times10^{-4}$ at 430 nm. With DDA, stability is maintained, reducing this rate to $(1\pm0.1)\times10^{-4}$ or lower. These represent preliminary results from small-scale laboratory synthesis, and there remains room for further optimization and improvement. Experimental results indicate an optimal DDA dosage range of n$_\mathrm{DDA}$:n$_\mathrm{Te}$ = (0.2$\sim$0.3):1. Two DDA incorporation methods were applied: addition during synthesis or during Te-loaded sample preparation. To date, both methods demonstrate comparable effectiveness in stabilizing Te-LAB solutions, though post-synthesis addition better maintains consistent DDA concentrations. Preliminary mechanistic studies on DDA's stabilizing effects have been conducted, with detailed results to be reported subsequently.

The resulting scintillators exhibit exceptional optical properties: the optimal Te-LAB sample demonstrates an attenuation length of 20.1 $\pm$ 1.1 m at 1\% Te loading. Long-term stability monitoring of 1\% and 3\% Te-LAB solutions confirms maintained spectral stability over approximately one year. However, the light yield is substantially affected by Te-loading; although it is comparable to that of the type I loading method reported in Ref. ~\cite{SNO+:2023}, it remains lower than the light yield of the Te-LS obtained via their type II loading method.

Currently, in addition to the Type I and Type II synthesis routes of the SNO+ collaboration~\cite{SNO+:2023}, we have developed two water-free synthesis routes for Te-diol compounds: the room-temperature method presented herein and the established high-temperature azeotropic distillation approach~\cite{IHEP:2023}. In both cases, a strict purification protocol is applied to all raw materials to meet rigorous purity standards. Therefore, the resulting Te-diol product is used directly for the preparation of Te-LS without further purification. At present, our experiments focus solely on the optical purity of the raw materials. While research on radiopurification is planned, it has not yet been conducted. The radiopurification of HD (b.p. 224\textdegree C) and DDA (b.p. 247\textdegree C) is proposed to be conducted via fractional distillation. This is a highly efficient and mature method for removing both radioactive impurities (e.g., thorium, uranium) and optical impurities from liquids. The principal challenge in removing radioactive impurities lies in the treatment of telluric acid, which is tackled through techniques such as self-scavenging as well as acid and thermal recrystallization~\cite{Hans:2015wba}, pH adjustment and coprecipitation/adsorption to ensure compliance with the stringent requirements of $0\nu\beta\beta$ experiments. 

Specifically, the light yield exhibits a noticeable reduction upon Te-loading, necessitating further investigation into enhancement strategies. Moreover, in contrast to the route reported in Ref.~\cite{SNO+:2023}, the use of MeOH poses potential safety risks for large-scale production. To address these issues, our future work will proceed in two directions: first, we aim to optimize the current process or explore safer, more efficient alternative synthesis routes; second, we will systematically examine structural modifications of Te-diol derivatives and their effects on optical properties, especially light yield. These efforts are directed toward developing advanced formulations that fully satisfy the demanding specifications for $0\nu\beta\beta$ detection.

\section*{Acknowledgment}

This work was supported in part by the National Natural Science Foundation of China (Grant Nos. 12141504 and 12125506), CAS Project for Young Scientists in Basic Research (Grant No. YSBR-099), Beijing Natural Science Foundation (Grant No. 1252035), and by the State Key Laboratory of Particle Detection and Electronics (Grant No. SKLPDE-KF-202403). The authors thank H.C. Han and E.Z. Zhang for the support in the initial stage of light yield measurement.

\end{document}